\documentclass{ws-rv9x6}
\usepackage{subfigure}     % required only when side-by-side / subfigures are used
\usepackage{ws-rv-van}     % numbered citation/references (default)
\makeindex
\begin{document}

\chapter[Using World Scientific's Review Volume Document Style]{Isoscalar and isovector neutron-proton pairing}\label{ra_ch1}

\author[A. V. Afanasjev]{A.\ V. Afanasjev}
%\index[aindx]{Author, F.} % or \aindx{Author, F.}
%\index[aindx]{Author, S.} % or \aindx{Author, S.}

\address{Department of Physics and Astronomy, Mississippi State
University, \\
MS 39762, USA, \\
afansjev@erc.msstate.edu}

\address{Joint Institute for Heavy-Ion Research, Oak Ridge, \\  
TN 37831, USA}

\begin{abstract}
 Neutron-proton ($np-$) pairing is expected to play an important role 
in the $N\approx Z$ nuclei. In general, it can have isovector and isoscalar 
character. The existence of isovector $np-$pairing is well established. 
On the contrary, it is still debated whether there is an isoscalar 
$np-$pairing. The review of the situation with these two types of pairing 
with special emphasis on the isoscalar one is presented. It is concluded 
that there are no substantial evidences for the existence of isoscalar 
$np$-pairing.
\end{abstract}

\body

%%%%%%%%%%%%%%%%%%%%%%%%%%%%%%%
\section{Introduction}
%%%%%%%%%%%%%%%%%%%%%%%%%%%%%%%

  The invent of new generation of detector facilities (such as
GAMMASPHERE and EUROBALL) and radioactive beams in the 90ies of
last century has opened up new avenues to study the nature of nuclear 
interactions, in particular, $np-$pairing at the $N = Z$ line. This 
also stimulated theoretical studies of this type of pairing.

  The existence of the $np-$pairing crucially depends upon the 
overlap between the neutron and proton wave functions.\footnote{It 
is frequently stated that near degeneracy of the proton and neutron 
Fermi surfaces favors the development of neutron-proton pairing. 
This is, however, not true considering that Coulomb interaction creates 
an energy gap of approximately 7 MeV between the proton and neutron 
states of the same structure (and respective Fermi surfaces). This 
fact is ignored in a number of publications.} Protons and neutrons 
occupy the same orbitals in $N=Z$ nuclei and this leads to an 
increased neutron-proton pair correlations which under specific 
circumstances can form $np-$pair condensate. A suppression 
of this type of pairing is expected if the system is driven out of 
the isospin-symmetric state. Thus, $np-$pairing is expected only at
$N=Z$ line or in its close vicinity \cite{WFS.71,SatW.97}. Indeed,
it is well known that in the nuclei away from the $N=Z$ line proton-proton 
($pp$) and neutron-neutron ($nn$) pairing dominate and there are no
signs of $np-$pairing. The mechanism driving this suppression is 
encountered not only in nuclei but also in other many-fermionic systems 
(such as superconductors and superfluids) where the particles lie on two 
different Fermi surfaces (see Ref.\  \cite{SL.00} for more details). 

  These $np-$correlations can be isoscalar and isovector. Figuring 
out whether they form a static pair condensate/pairing (an average field) 
in respective channel has been a challenge since medium mass  $N=Z$ nuclei 
have come into reach of experiment. In this manuscript, I review the situation 
with the current understanding of isoscalar and isovector $np-$pairing. A 
specific attention is paid to isoscalar $np-$pairing since it is not clear 
at present whether this type of pairing exists or not. The general consideration 
of the $np-$pairing is presented in Sect.\ \ref{Theory}. The impact of the 
$np-$pairing on different physical observables and processes in non-rotating 
and rotating nuclei is discussed in Sects.\ \ref{Norotation} and \ref{np-rot}, 
respectively.

%%%%%%%%%%%%%%%%%%%%%%%%%%%%%%%%%%%%%%%%%%%%%%%%%%%%%%%%%%%%%
\section{Neutron-proton pairing: general considerations}
\label{Theory}
%%%%%%%%%%%%%%%%%%%%%%%%%%%%%%%%%%%%%%%%%%%%%%%%%%%%%%%%%%%%%

  Isotopic invariance of nucleon-nucleon interaction tells
us than the nuclear components of the interaction in the systems 
proton-proton, neutron-neutron and neutron-proton are very similar. 
A nucleon with isospin  quantum number $\tau=1/2$ may be in one 
of two states,  $\tau_z=-1/2$ (proton) and  $\tau_z=+1/2$ (neutron).
Nuclear many-body states are labeled with isospin quantum number 
$T$, whose third component is its projection $T_z=(N-Z)/2$
($N$ and $Z$ are neutron and proton numbers of the nucleus, 
respectively).

  Let me consider a pair of two nucleons. For such a system, two 
distinct isospin states with $T=1$ and $T=0$ can be defined. The spin 
projections $T_z=-1,0,1$ are possible for a $T=1$ nucleon-nucleon 
system. Here $T_z=-1$ corresponds to a proton-proton system, $T_z=1$ 
to a neutron-neutron system, and $T_z=0$ to a neutron-proton system. 
The nucleons in the $T=1$ system have total spin $J=0$ in order to 
ensure antisymmetry of the total nucleon-nucleon wave function. For 
the same reasons, $T=0$ proton-neutron systems can have only $T_z=0$;
the situation with total spin is discussed below. 

  The scattering of the nucleon pairs with given quantum numbers
of isospin $t$\footnote{The lower-case letter $t$ is used for the 
isospin of the pair-field in order to avoid the confusion with the 
total isospin of the states denoted by $T$.} and angular momentum 
$J$ is responsible for different kinds of pairing correlations 
\cite{G.79,DH.03}. The pair potential $\Delta_{Jt}$ is also defined by 
the spin and angular momentum of pair. It is well known that in 
even-even nuclei isovector $t=1$ like-particle pairing is responsible 
for the spins and parities ($J^{\pi}=0^+$) of the ground states and 
for appreciable separation in energy of ground and excited states. For 
this pairing, a nucleon pair couples to angular momentum $J=0$.

\begin{figure}[th]
\centerline{\psfig{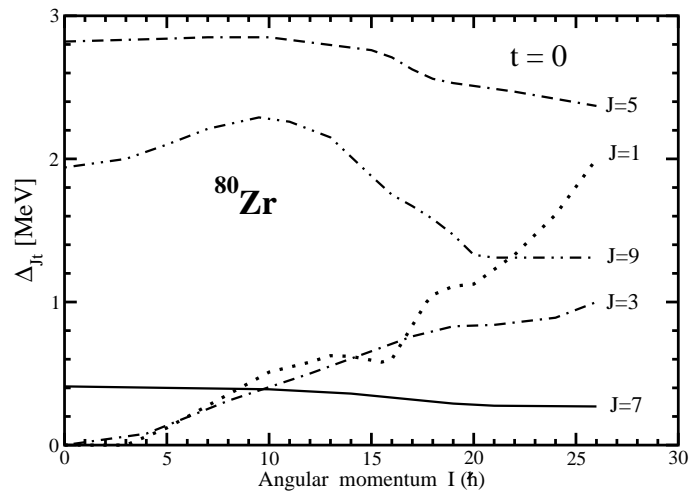}}
\caption{Angular momentum components of the pair potential 
$\Delta_{J,t=0}$ for the $t=0$ pair band in $^{80}$Zr. Based
on the results presented in Fig.\ 9 of Ref.\cite{G.01}.
\label{Pair-pot}}
\end{figure}

 The situation is different for neutron-proton pairing. There 
are two possible types of pairing: isovector one with $t=1$ and $J=0$ and 
isoscalar one with $t=0$. It is frequently stated that in the case 
of isoscalar pairing the dominant components of pair potential 
correspond to either $J=1$, or $J=J_{max}=2j$, where $j$ is the
nucleon angular momentum. However, the results of the calculations
of Ref.\ \cite{G.01} presented in Fig.\ \ref{Pair-pot} show that 
this is not always a case. Indeed, at spin $I=0$ in the $t=0$ pair 
band of $^{80}$Zr there is no $J=1$ or $J=3$ pairs. The pair potential 
is dominated by the $J=5$ pairs and $J_{max}=2j=9$ pair comes 
only as a second in importance.

%%%%%%%%%%%%%%%%%%%%%%%%%%%%%%%%%%%%%%%%%%%%%%%%%%%%%%%%%%%%%%%%%%%%%%%%%%%%%%%%
%%%%%%%%%%%%%%%%%%%%%%%%%%%%%%%%%%%%%%%%%%%%%%%%%%%%%%%%%%%%%%%%%%%%%%%%%%%%%%%%%%%%

  Earlier calculations have pointed on the exclusiveness of the 
$t=0$ and $t=1$ $np-$pairing phases \cite{GSG.68,GSBG.70}. 
However, more recent calculations show that $t=0$ and $t=1$ pairing
phases can coexist. This was shown in Ref.\ \cite{G.99} within the 
isospin generalized BCS and HFB frameworks based on the G-matrix 
interaction. Ref.\cite{SatW.00} illustrated that the sudden 
phase transition between the $t=0$ and $t=1$ pairing modes becomes 
smeared out in number-projected Lipkin-Nogami (LN) calculations.

 When considering $np-$pairing it is important to remember the basic 
difference between shell model and mean field (MF)/density 
functional (DFT) models since the neglect of this difference 
frequently leads to confusions and contradictions. The shell model 
Hamiltonian is usually written in the particle-particle representation.
Thus, in the shell model there is no distinct division into pairing- and
single-particle (mean) fields.  On the contrary, the configuration space 
of the MF and DFT models is separated into particle-hole (mean field) 
and particle-particle 
(pairing field) channels. As a consequence, the shell model definition of 
pairing in terms of  $L=0, 
S=1, T=1$ and $L=0, S=1, T=0$ pairs is completely inappropriate from 
the point of MF/DFT models (see discussion in Ref.\ \cite{SatW.00} and 
references therein). This means that the existence of 
isoscalar and isovector $np-$pair correlations in spherical shell model 
is not equivalent to the existence of isoscalar and isovector $np-$pairing 
[pair condensate] in the MF/DFT frameworks. As a consequence, I only
consider here the results obtained in the MF/DFT frameworks.

%%%%%%%%%%%%%%%%%%%%%%%%%%%%%%%%%%%%%%%%%%%%%%%
\subsection{Isovector neutron-proton pairing}
\label{isnpp}
%%%%%%%%%%%%%%%%%%%%%%%%%%%%%%%%%%%%%%%%%%%%%%%

 At present, the situation with the isovector $np$-pairing is most 
clarified. The isovector $np-$pairing is absolutely necessary in order 
to restore the isospin symmetry of the total wave function 
\cite{FS.99-NP}. Its strength is well defined by the isospin symmetry.  
A number of experimental 
observables such as binding energies of the $T=0$ and $T=1$ states in 
even-even and odd-odd $N=Z$ nuclei \protect\cite{MFC.00,V.00,Rb74},
the observation of only one even-spin $T=0$ band in $^{74}$Rb \cite{Rb74} 
instead of two nearly degenerate bands expected in the case of no $t=1$ 
$np$-pairing {\it clearly point on the existence of pair condensate 
in this channel}. The  analysis of pairing vibrations around $^{56}$Ni 
indicates  a  collective behavior of the isovector pairing vibrations 
but does not support any appreciable collectivity in the isoscalar 
channel \cite{bes-rev,MFCC.00}. The detailed discussion of binding energies 
of the $T=0$ and $T=1$ states in even-even and odd-odd $N=Z$ nuclei as 
well as pairing vibrations around $^{56}$Ni is given in the contribution 
of A.\ Macchiavelli in this Volume \cite{M.12}.

%%%%%%%%%%%%%%%%%%%%%%%%%%%%%%%%%%%%%%%%%%%%%%%
\subsection{Isoscalar neutron-proton pairing}
%%%%%%%%%%%%%%%%%%%%%%%%%%%%%%%%%%%%%%%%%%%%%%%

   While the situation with isovector $np-$pairing is settled, the one with 
isoscalar $np-$pairing is full of controversies. These controversies are 
generally related to the microscopic origin of isoscalar $np-$pairing and 
whether the isoscalar $np-$pair correlations lead to a pair condensate.

  The calculations with the realistic (bare) forces (Paris force, Argonne V14 
force) indicate that the isoscalar pairing gap in the symmetric nuclear matter is 
3 times larger than the isovector one \cite{GSMLSS.01}. In finite nuclei with 
$Z=N=35$, calculated isoscalar pairing gap is of the order of 3 MeV \cite{GSMLSS.01},  
while the experimental isovector pairing gap is around 1.8 MeV (see Fig.\ 4 in Ref.\  
\cite{MFC.00}). However, despite that no convincing fingerprints of isoscalar 
$np$-pairing has been found so far (see discussion below).

  The potential problem is due to the transition from realistic to effective 
interaction: the extremely strong $t=0$ $np-$pairing emerges essentially from the fact 
that with respect to the $t=1$ channel, dominated by the central force, the tensor 
force is acting additionally. However, the medium modification (screening) of the 
tensor force is still controversial subject \cite{ZZ.91}. For example, higher shell 
admixtures make the tensor force appear weaker in the valence space \cite{FZC.97}. 
In addition, one cannot exclude the possibility that the tensor force is largely 
screened in the medium, and, thus, the enhancement of the T=0 gap values may be brought 
back closer to the values of the $T=1$ case \cite{GSMS.99}. 

 While the structure of interaction (central force) is the same in isovector 
pairing channel of the theories based on realistic and effective forces, the 
addition of tensor component into isoscalar pairing channel of the models based 
on effective forces may be necessary for a correct description of $np$-pairing 
in this channel. In the existing mean-field models, the tensor component of
pairing is neglected. Although, some attempts were made to approximate bare
tensor interaction by effective density dependent zero-range $\delta$-force
\cite{GSMS.99,GSMLSS.01}, the validity of such an approximation for different
physical observables has not been tested in the mean field calculations.

  Recent Hartree-Fock-Bogoliubov (HFB) studies \cite{DMS.10} for finite nuclei 
with chiral N$^3$LO two-nucleon interaction for pairing led to the results 
which are opposite to the ones discussed above. They showed that this type of 
nuclear forces favors isovector over isoscalar pairing, except in low-$j$ orbitals. 
The supression of isoscalar pairing has been traced to the effects of spin-orbit 
splitting, the $D$ waves and additional repulsive $^1P_1$ channel. Note that
the role of spin-orbit field in the suppression of isoscalar pairing has also
been discussed in Refs.\ \cite{PP.98,BL.10}.

  The presence or absence of isoscalar $np$-pair condensate sensitively
depends on the strength of the pairing in this channel (see Sect.\ 
\ref{Wigner-sect} below). At present, it is obvious that microscopic 
theories give no clear guidance on what strength has to be used for isoscalar 
$np-$pairing in the MF/DFT models. It was suggested to extract the strengths 
of the $t=0$ $np-$pairing from experimental Wigner energies (see Sect.\ 
\ref{Wigner-sect} below). However, there are alternative explanations of the 
Wigner energy which do not involve $t=0$ $np-$pairing. As a consequence, on 
the MF/DFT level there is no generally accepted procedure on how to extract 
the strength of isoscalar $np-$pairing. This situation is clearly unsatisfactory. 
Thus, the systematic comparison betwen theory and experiment with the goal to 
find the evidences for isoscalar $np$-pairing and physical observables 
sensitive to it becomes imperative. Such a comparison is presented below.

%%%%%%%%%%%%%%%%%%%%%%%%%%%%%%%%%%%%%%%%%%%%%%%%%%%%%%%%%%%%%%%%%%%%%%%%%%%%%%%%
\begin{figure}[th]
\centerline{\psfig{file=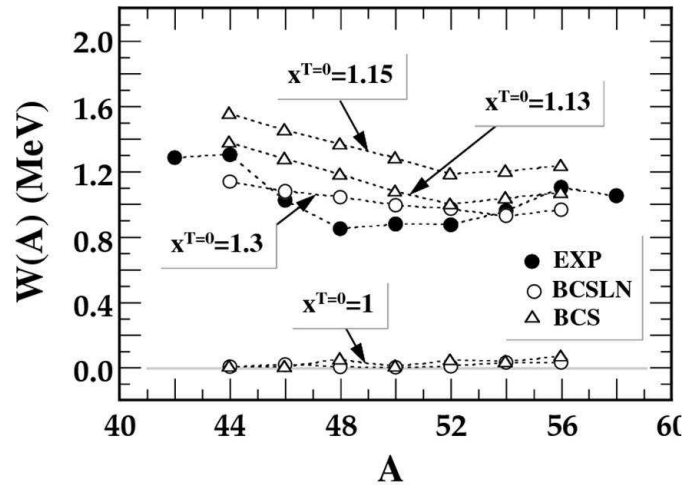,width=9.0cm}}
\caption{Experimental and calculated strength $W(A)$ of the Wigner energy 
for $pf-$shell nuclei. The results of calculations for different values 
of $x_0^{t=0}$ and different models are shown. From Ref.\ \cite{SatW.00}.
Note that the authors of this reference use capital letter $T$ for the
isospin of the pair-field, while lower-case $t$ is used for it in the current
manuscript.
\label{Wig}}
\end{figure}
%%%%%%%%%%%%%%%%%%%%%%%%%%%%%%%%%%%%%%%%%%%%%%%%%%%%%%%%%%%%%%%%%%%%%%%%%%%%%%%%%%%%

%%%%%%%%%%%%%%%%%%%%%%%%%%%%%%%%%%%%%%%%%%%%%%%%%
\section{Neutron-proton pairing at no rotation}
\label{Norotation}
%%%%%%%%%%%%%%%%%%%%%%%%%%%%%%%%%%%%%%%%%%%%%%%%%

%%%%%%%%%%%%%%%%%%%%%%%%%%%%%%%%%%%%%%%%%%%%%%%%%%%%%%%
\subsection{Wigner energy}
\label{Wigner-sect}
%%%%%%%%%%%%%%%%%%%%%%%%%%%%%%%%%%%%%%%%%%%%%%%%%%%%%%%
 
 It is well know that a term proportional to isospin T has to 
be included into nuclear mass formulae in order to reproduce the 
isospin dependence of masses \cite{MS.82}. This term called Wigner 
energy has a form $E_W=W(A)|N-Z|/A$ in which $W(A)$ stands for 
mass-dependent strength. It gives rise to a cusp at $N=Z$ in 
the curves of masses along an isobaric chain. The physical 
origin of this energy has not definitely been  established untill 
now and it still remains the subject of the debate (see Sect. II 
in Ref.\cite{N.09} for a recent review).
The modern mean field models or DFT do not explain 
it; this term is added as an ad-hoc phenomenological term. 

  As suggested in Ref.\ \cite{SatW.97} one of possible microscopic 
explanations of this term involves the isoscalar $t=0$ $np-$pairing. 
The experimental Wigner energies can be reproduced in this scenario
(Ref.\ \cite{SatW.00}) but this requires the strength of isoscalar 
($G_{np}^{t=0}$) $np-$pairing which is larger than the one of 
isovector ($G_{np}^{t=1}$) $np-$pairing. No isoscalar $np-$pair 
condensate is 
formed for the case $G_{np}^{t=0}=G_{np}^{t=1}$ (see Fig.\ \ref{Wig}).
One then can define the scaling factor $x_{0}^{t=0}=G_{np}^{t=0}/G_{np}^{t=1}$.
The fit to experimental Wigner energies gives the $x_0^{t=0}$ values
of $\sim 1.13$ and $\sim 1.30$ for BCS and BCSLN models in the $fp$ 
shell (see Fig.\ \ref{Wig}) and $\sim 1.25$ in the BCSLN model in the 
$A\sim 76$ mass region. These high values of $x_0^{t=0}$ lead to a visible 
impact of the $t=0$ $np-$pairing on the rotational properties of the 
$N\approx Z$ nuclei at high spin \cite{SatW.97,SatW.00}. However, their 
detailed analysis discussed in Sect.\ \ref{np-rot} does not support the 
presence of isoscalar $np-$pairing. 

 Alternative explanations of the Wigner energy which do not involve 
isoscalar $np-$pairing have been proposed in Refs.\cite{N.02,N.03,BF.12}.
It was suggested in Refs.\cite{N.02,N.03}  that the RPA correlation energy should be 
taken into account in order to describe experimental masses in the vicinity 
of the $N\approx Z$ line. In this formalism, the Wigner energy results from the 
collectivity of the isorotation, which itself is the result of the isorotational 
noninvariance of the isovector pair field. In another scenario \cite{BF.12}, 
the combination of an isorotational invariant effective interaction in the 
particle-hole channel with isovector pairing interaction gives the Wigner energy, 
provided the pairing correlations are treated beyond mean field approximation 
and isospin is conserved.

  One should note, however, that all of these results have to be taken with 
a grain of salt because they are probably crude approximations to the real 
situation due to employed simplifications. For example, Refs.\ \cite{SatW.00} 
ignore the conservation  of isospin and correlations beyond mean field, 
whereas the results of Refs.\ \cite{N.02,N.03} were obtained in schematic 
model.
% and the model of Ref.\ \cite{BF.12} is restricted to monopole
% pairing. The later is probably not that bad approximation, but one has to 
% check how the results will be affected by the inclusion of quadrupole
% pairing. 

%%%%%%%%%%%%%%%%%%%%%%%%%%%%%%%%%%%%%%%%%%%%%%%%%%%%%%%%%%%%%%
\subsection{Binding energies of the  $T=0$ and $T=1$ states 
in even-even and odd-odd $N=Z$ nuclei}
%%%%%%%%%%%%%%%%%%%%%%%%%%%%%%%%%%%%%%%%%%%%%%%%%%%%%%%%%%%%%%

 The analysis of experimental binding energies of the $T=0$ and $T=1$ 
states in even-even and odd-odd $N=Z$ nuclei \protect\cite{MFC.00,V.00,Rb74} 
clearly points on the existence of pair condensate in the isovector 
channel but provides no evidence for an isoscalar pair condensate in such 
nuclei. The detailed discussion of this topic is given in the contribution 
of A.\ Macchiavelli in this Volume \cite{M.12}. The observed spectra of
adjacent even-even and odd-odd nuclei $N=Z$ nuclei are distinctly 
different. This also allows to exclude pure $t=0$ $np-$pairing field with 
$\Delta$ larger than the single-particle level distance \cite{FS.00PS}.

%%%%%%%%%%%%%%%%%%%%%%%%%%%%%%%%%%%%%%%%%%%%%%%%%%%%%%%
\subsection{Neutron-proton pairing in transfer reactions}
%%%%%%%%%%%%%%%%%%%%%%%%%%%%%%%%%%%%%%%%%%%%%%%%%%%%%%%

The collectivity of $np-$pairing correlations can be accessed by 
means of pair transfer from the $T=1(0)$ ground state of the $A+2$ 
$(N=Z)$ nucleus to the ground state of the $A$ $(N=Z)$ nucleus. 
The analysis of the influence of the $np-$pairing on $np-$pair transfer 
in $N=Z$ nuclei within a single-$j$ shell model space with allowance 
for both $t=0$ and $t=1$ pairing interactions\cite{F.70,F.71} lead to 
the conclusion that $np-$pairing can enhance the cross-section by a 
factor 3 as compared to conventional shell-model calculations. However, 
more sophisticated analysis\cite{GSW.04} pointed out that the fundamental 
difference in the structure between the $t=0$ vacua in even-even and 
odd-odd nuclei results in a quenching of the $T=0$ pair transfer even 
in the presence of strong $t=0$ $np-$pairing. So far experimental 
measurements of the $np-$ pair transfers in the $N=Z$ nuclei have not 
provided conclusive answer on whether the $t=0$ $np$-pair 
condensate is formed \cite{M.12}.

%%%%%%%%%%%%%%%%%%%%%%%%%%%%%%%%%%%%%%%%%%%%%%%%%%%%%%%%
\subsection{Pairing vibrations}
%%%%%%%%%%%%%%%%%%%%%%%%%%%%%%%%%%%%%%%%%%%%%%%%%%%%%%%%

 Near closed shells, the strength of the pairing force relative to the 
single-particle level spacing is expected to be less than the critical 
value needed to obtain a superconducting solution, and the pairing field 
then gives rise to a collective phonon \cite{bes-rev}. It then seems
natural to ask whether $t=0$ collective effects may show up as a vibrational
phonon? A detailed analysis of isovector pairing vibrations around
$^{56}$Ni presented in Refs.\ \cite{bes-rev,MFCC.00} confirms their 
collectivity. On the contrary, the analysis of the excitation spectrum 
around this nucleus indicates only a single-particle character for the 
isoscalar channel \cite{MFCC.00}.

%%%%%%%%%%%%%%%%%%%%%%%%%%%%%%%%%%%%%%%%%%%%%%%%%%%%%%%%%
%\subsection{Other possible physical observables}
%%%%%%%%%%%%%%%%%%%%%%%%%%%%%%%%%%%%%%%%%%%%%%%%%%%%%%%%%
%
% Isoscalar $np-$pairing may play a role in single-$\beta$ decay
% [2,17] (see, however, Refs. [18,19]), double $\beta$ decay [20,21],
% $\alpha$- decay,
% and $alpha$-correlations [24,25]. However, because no 
% symmetry unrestricted mean-field (and beyond mean field) calculations
% of $np-$pairing, based on realistic effective interaction and the
% isospin-conserving formalism, have been carried out so far, no
% hard evidence for the elusive t = 0 np pairing phase has yet
% been found.

%%%%%%%%%%%%%%%%%%%%%%%%%%%%%%%%%%%%%%%%%%%%%%%%%%%%%%%%%%%%%%%%%%%%%%%%%%%%%%%%
\begin{figure}[th]
\centerline{\psfig{file=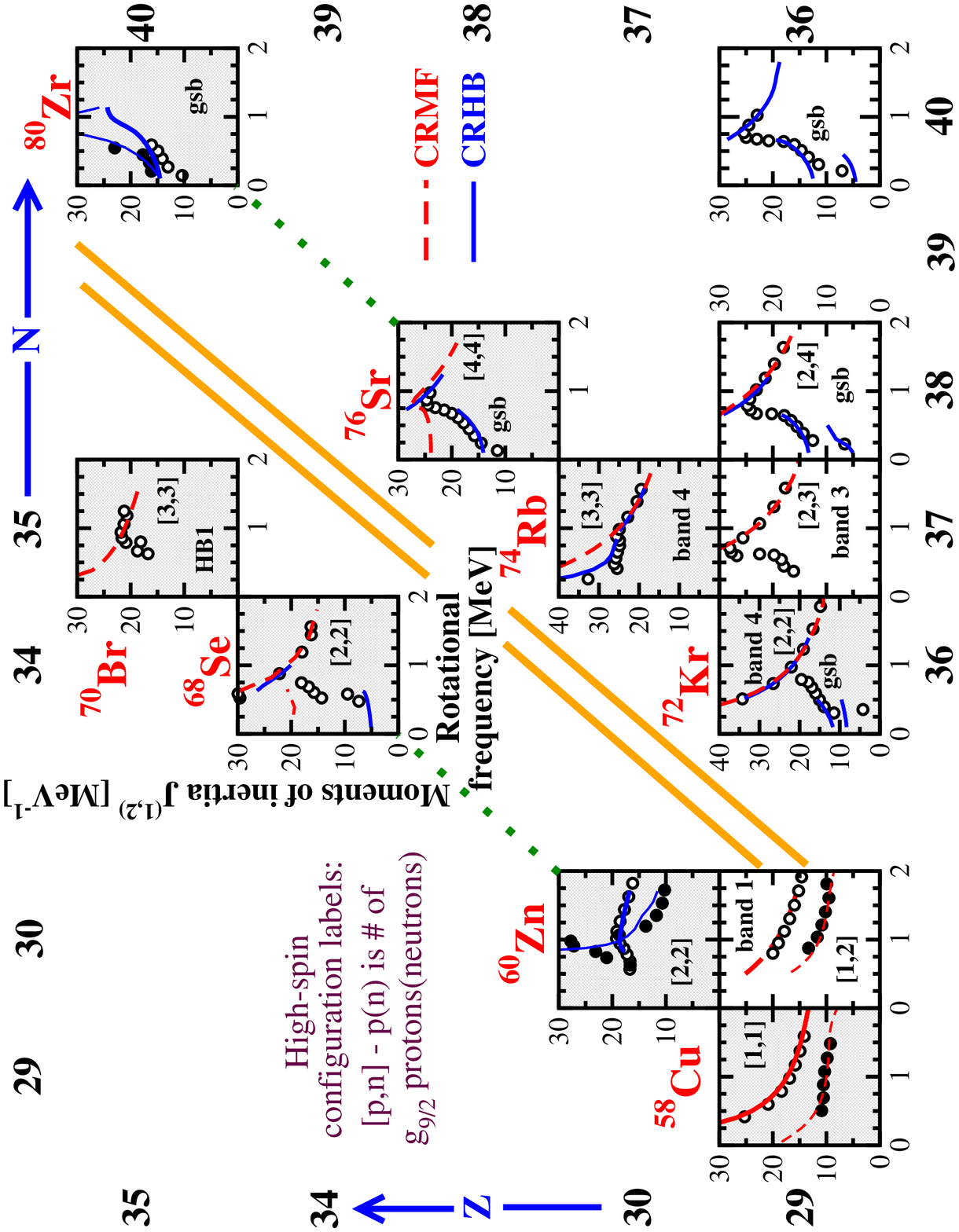,width=11cm}}
\vspace{-0.8cm}
\caption{The kinematic moments of inertia $J^{(1)}$ of rotational structures 
in the $N\approx Z$ nuclei compared with the results of the CRMF and 
CRHB calculations. The shaded background is used for $N=Z$ nuclei. The 
vertical scale of the panels for $^{72}$Kr and $^{74}$Rb is different 
from the one of the other panels. The figure is based on the results 
published in Refs.\ 
\protect\cite{Rb74,AF.05,Br70,A60,74Kr,Kr73,72Kr,Cu59,Sr76}. Note that in
few cases the results for dynamic moments of inertia $J^{(2)}$ are shown. In these 
cases, thick and thin lines are used for calculated kinematic and dynamic 
moments of inertia, respectively. Experimental kinematic and dynamic moments 
of inertia are shown by open and solid circles, respectively. The results
of the CRHB calculations at low spin are shown both for prolate and 
oblate minima in few cases; in a given nucleus calculated $J^{(1)}$ in oblate 
minimum is lower than the one in prolate minimum.
\label{syst}}
\end{figure}
%%%%%%%%%%%%%%%%%%%%%%%%%%%%%%%%%%%%%%%%%%%%%%%%%%%%%%%%%%%%%%%%%%%%%%%%%%%%

%%%%%%%%%%%%%%%%%%%%%%%%%%%%%%%%%%%%%%%%%%%%%%%%%%%%%
\section{Neutron-proton pairing in rotating nuclei}
\label{np-rot}
%%%%%%%%%%%%%%%%%%%%%%%%%%%%%%%%%%%%%%%%%%%%%%%%%%%%%

 The properties of the $N\approx Z$ rotating nuclei were in the focus 
of the debate on the existence of isoscalar and isovector $np$-pairing. The
following physical observables
\begin{itemize}
\item
the size of the moments of inertia \cite{SatW.97,SatW.00,G.01},

\item
the frequencies at which the pairs of particles align their angular momentum 
(band crossing frequencies\footnote{Note that different authors attribute
the shift of crossing frequency in rotational bands either to isovector 
\cite{FS.99-NP,FS.99,KZ.98} or isoscalar \cite{SW.00,kr72t0,KZ.98}
$np-$pairing or their combination \cite{SatW.97,SatW.00,G.01}.}
\cite{KZ.98,FS.99-NP,FS.99,SW.00,kr72t0,SatW.00,G.01}),

\item
deformation properties \cite{TWH.98},

\item
unexpected mixing of configurations \cite{Br70,Kr73,AF.05},

\item
the properties of terminating states \cite{TWH.98,SS.06}

\end{itemize}
have been discussed in the literature as possible indicators of the 
$np$-pairing in rotating $N\approx Z$ nuclei.

 As discussed in Sect.\ \ref{isnpp} the evidences for the existence of 
isovector $np-$pairing are very strong. The investigation of rotational
structures, namely, the observation of only one even-spin $T=0$ band in 
$^{74}$Rb \cite{Rb74} instead of two nearly degenerate bands expected in 
the case of no $t=1$ $np$-pairing supports the existence of pair 
condensate in this channel.

 On the other hand, no such strong arguments exist for isoscalar $np-$pairing. 
Thus, it was suggested in Ref.\ \cite{AF.05} to investigate rotating $N\approx Z$ 
systems within the isovector mean-field theory \cite{FS.99-NP} with the goal to 
see whether the discrepancies between this theory and experiment can be related to 
$t=0$ $np-$pairing. This theory assumes that there is no isoscalar $np-$pairing, 
but takes into account isovector $np-$pairing and isospin symmetry conservation. 
A clear advantage of this theory  is the fact that standard mean field models with 
only $t=1$ like-particle pairing can be employed. The basis modification of these 
theories lies in adding the isorotational energy term $T(T+1)/2{\cal J}_{iso}$ 
to the total energy. Since, however, all low-lying rotational bands in even-even 
$N=Z$ nuclei have isospin $T=0$, this term vanishes. On the level of accuracy of the 
standard mean-field calculations, the restoration of the isospin symmetry (which 
takes care of the t=1 $np$ pair field) changes only the energy of the $T=1$ states 
relative to the $T=0$ states \cite{FS.99-NP}. With this in mind, the rotating 
properties were studied by means of the cranked Relativistic Hartree-Bogoliubov 
\cite{CRHB} (CRHB) theory.

  At high spin, the impact of $t=1$ pairing is negligible and consequently it can be 
neglected. In such situation, the isospin broken at low spin by isovector pairing is 
conserved automatically \cite{G.99}. Thus, the high spin ($I\geq 15\hbar$) states 
were systematically studied by the cranked Relativistic Mean Field (CRMF) 
\cite{VRAL.05} approach which assumes zero pairing\footnote{In addition, 
the cranked Nilsson-Strutinsky (CNS) approach \cite{PhysRep} has been used for the 
study of high spin states.  Note that the results of the CNS calculations are 
similar to the CRMF ones so they are not discussed here.}. In the calculations without 
pairing, the shorthand notation $[p,n]$ indicating the number $p(n)$ of occupied 
$g_{9/2}$ proton (neutron) orbitals is used for labeling of the configurations.

%%%%%%%%%%%%%%%%%%%%%%%%%%%%%%%%%%%%%%%%
\subsection{Moments of inertia}
%%%%%%%%%%%%%%%%%%%%%%%%%%%%%%%%%%%%%%%%

 Since $t=0$ pairs carry angular momentum, a $t=0$ $np-$pair field is expected 
to increase the moments of inertia \cite{SatW.97,SatW.00,G.01}.  In contrast 
to the {\it static} $t=1$ pair field, which is suppressed by the Coriolis 
anti-pairing (CAP) effect, {\it static} $t=0$ $np-$pairing is favored by rotation. 
The suggested microscopic mechanism behind that is the following \cite{SatW.97}. 
The rotation increases the number of pairs of nucleons with parallel coupled 
angular momenta, thus enforcing the $t=0$ $np-$pairing. In this pairing phase, 
angular momentum is built by the  $np-$pairs smoothly aligning along the 
rotational axis, without involving any pair breaking mechanism typical for $t=1$ 
pairing. Note that $t=0$ $np-$pairing saturates with increasing frequency. Thus, 
at large angular momentum, where the {\it static} $t=1$ field is destroyed, a 
substantial difference between experimental moments of inertia and the ones 
obtained in the calculations without pairing may indicate the presence of the 
$t=0$ $np-$pair field.

  Fig.\ \ref{syst} shows that the moments of inertia of rotational bands in the 
$N\approx Z$ nuclei are well reproduced by the CRHB calculations before first 
band crossings. The accuracy is the same as for neighboring $N \neq Z$ nuclei. The 
CRHB calculations as well as the ones of Ref.\ \cite{SatW.00} indicate that after 
first proton and neutron paired band crossings the static $t=1$ pairing correlations 
are essentially gone. Indeed, above these crossings the moments of inertia obtained
in the CRMF and CRHB calculations are very similar. The experimental moments of inertia 
of the $N\approx Z$ nuclei above band crossings are well reproduced by the unpaired 
CRMF calculations (as well as cranked Nilsson-Strutinsky calculations \cite{AF.05}),  
where it turned out to be important that the response of the nuclear shape to rotation 
was properly taken into account. {\it Thus, no systematic underestimate of the moments 
of inertia, which could be taken as an evidence for a $t=0$ $np$-pair field, could 
be identified.}

% In this regime of fast rotation, the 
% calculations without pairing provide very good description of the rotational 
% properties of different types of bands (terminating, superdeformed etc.) in 
% different regions of the periodic table (see Refs.\ \cite{PhysRep,A150,A60}).
% In accordance with this general observation, the experimental  moments of 
% inertia in the $N=Z$ (see present manuscript for $^{68}$Se, $^{72}$Kr,
% $^{76}$Se and $^{80}$Zr nuclei and Refs.\ \cite{Rb74,Zn60SD,A60,Br70} for
% $^{58}$Cu, $^{60}$Zn, $^{70}$Br and $^{74}$Rb nuclei), $N=Z+1$ 
% (see Refs.\ \cite{73Kr} [$^{73}$Kr] and \cite{Cu59} [$^{59}$Cu])
% and $N=Z+2$ (see Sec.\ \ref{74Kr-unpair} [$^{74}$Kr] and
% Ref.\ \cite{A60} [$^{62}$Zn]) 

%%%%%%%%%%%%%%%%%%%%%%%%%%%%%%%%%%%%%%%%%%%%%%%
\subsection{Band crossing frequencies}
%%%%%%%%%%%%%%%%%%%%%%%%%%%%%%%%%%%%%%%%%%%%%%%
 
  A delay of the first band crossing in the ground-state band of an 
even-even $N = Z$ system has been discussed as an evidence for $t = 0$ 
$np-$pairing in Refs.\ \cite{SW.00,kr72t0}. HFB \cite{SW.00} and 
cranked shell model \cite{kr72t0} calculations  in the $f_{7/2}$ subshell 
at fixed deformation indicate that the increase of the value of 
the $t = 0$ $np-$pair strength results in a delay of the crossing 
frequency in the ground-state band of $N = Z$ even-even nuclei.

  However, cranked shell model investigations \cite{FS.99-NP,KZ.98,FS.99} 
at fixed deformation show that such a delay can also be caused by the 
$t = 1$ $np-$pairing. On the contrary, more realistic total routhian 
surface calculations (TRS) with approximate particle number projection by 
means of the Lipkin-Nogami method show that in the case of superdeformed 
band in $N=Z$ $^{88}$Ru nucleus the paired band crossing takes place 
earlier if the isoscalar $np-$pairing is present (see Fig. 9 in Ref.
\cite{SatW.00}).

  Most of these investigations ignore the isospin conservation
\cite{FS.99-NP,G.99} and deformation changes \cite{AF.05} that 
are expected to play a crucial role in the $N\sim Z$ nuclei. Consequently, 
at present there are no reliable theoretical predictions on the magnitude 
of the shift (if any) of the band crossing frequencies in the $N = Z$ 
nuclei as compared with the $N \neq Z$ nuclei.
 
  The CRHB calculations within the framework of isovector mean
field theory provide rather good description of band crossings in the
$N\sim Z$ (see Fig.\ \ref{syst} and detailed discussion in Refs.\ \cite
{AF.05,Sr76}) which is comparable with the one achieved in the nuclei 
away from the $N=Z$ line. Similar level of agreement is achieved also
in the TRS \cite{WS.01} and projected shell model \cite{S.04} calculations 
without $np-$pairing. These results substantially weaken the argumentation 
if favor of the presence of the $t = 0$ $np-$pairing.

%%%%%%%%%%%%%%%%%%%%%%%%%%%%%%%%%%%%%%%%%%%%%%%
\subsection{Deformation properties}
%%%%%%%%%%%%%%%%%%%%%%%%%%%%%%%%%%%%%%%%%%%%%%%

\begin{figure}[th]
\centerline{\psfig{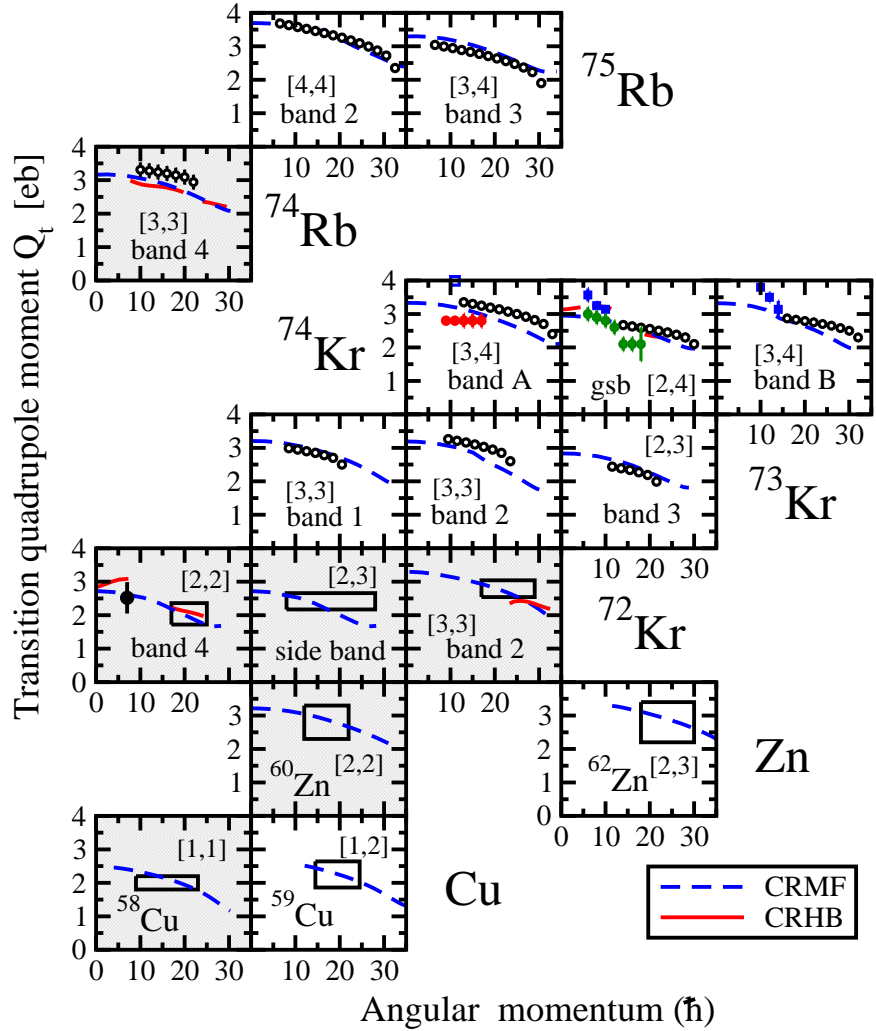}}
%\vspace*{8pt}
\caption{Transition quadrupole moments as a function of angular momentum. 
Experimental data are displayed either by data points (when available with
most recent ones shown by open circles) or by boxes. The boxes display the 
measured transition quadrupole moments and their uncertainties within the 
measured spin range. The results of the CRMF and CRHB calculations are 
shown. The shaded background is used for $N=Z$ nuclei. Experimental data 
and the results of calculations are taken from Ref.\ \cite{A60}($^{58}$Cu, 
$^{60,62}$Zn), Ref.\ \cite{Cu59}($^{59}$Cu), Ref.\ \cite{72Kr}($^{72}$Kr), 
Ref.\ \cite{73Kr74Rb-Qt}($^{73}$Kr), Refs.\ \cite{74Kr,74Kr-low}($^{74}$Kr), 
Ref.\ \cite{73Kr74Rb-Qt}($^{74}$Rb), and Ref.\ \cite{75Rb-nt}($^{75}$Rb).
\label{qt-sys}}
\end{figure}

 It was predicted in Ref.\ \cite{TWH.98} that the $t=0$ $np$-pairing 
generates an enhancement of the quadrupole deformation in the $N=Z$ 
nuclei. Fig.\ \ref{qt-sys} compares all availabe measured transition 
quadrupole moments $Q_t$ of observed bands in the $N\approx Z$ $A=58-75$
nuclei with the 
ones of assigned configurations. These data (both absolute values and 
relative changes in $Q_t$ with particle number and spin) agree rather 
well with the results of the CRMF, CRHB and CNS calculations (see Refs.\ 
\cite{Cu59,A60,AF.05,72Kr,73Kr74Rb-Qt,74Kr,74Kr-low,75Rb-nt} for more 
detailed discussion). One can also see that subsequent additions of $g_{9/2}$ 
particle(s) increase the transition quadrupole moment both in calculations 
and experiment. This analysis indicates that {\it no enhancement of 
quadrupole deformation in the $N=Z$ nuclei (which is expected in the 
presence of $t=0$ $np$-pairing \cite{TWH.98}) is required in order to 
reproduce experiment within the framework of isovector mean field 
theory.}

%%%%%%%%%%%%%%%%%%%%%%%%%%%%%%%%%%%%%%%%%%%%%%%%%%%
\subsection{Unexpected mixing of configurations}
%%%%%%%%%%%%%%%%%%%%%%%%%%%%%%%%%%%%%%%%%%%%%%%%%%%

In some nuclei, the [2,2] and [3,3] configurations are located very close 
in energy (see Fig.\ 14 in Ref.\ \cite{Br70} for $^{70}$Br, Fig.\ 10 in Ref.\ 
\cite{AF.05} for $^{72}$Kr and Fig.\ 6 in Ref.\ \cite{Kr73} for $^{73}$Kr). 
If the $t=0$ $np$-pairing is present, then these configurations are 
expected to be mixed.  A mixing represents the scattering of a proton and 
neutron on identical negative parity $N=3$ orbitals into identical $g_{9/2}$ 
orbitals, and vise versa. Such pair has an isospin $t=0$ since the proton and 
neutron are in the same space-spin state. Although some indications of
a mixing in these configurations exist (especially in $^{73}$Kr \cite{Kr73}), 
it does not provide a sufficient evidence for the presence of a $t=0$ pair 
field (see detailed discussion in Refs.\ \cite{Br70,Kr73,AF.05}). Rather it 
may indicate weak dynamical $t=0$ pair correlations as suggested by the Monte 
Carlo shell model calculations \cite{DKLR.97,AF.05} or just mixing of 
energetically close configurations by residual interaction \cite{Kr73,AF.05}. 

%%%%%%%%%%%%%%%%%%%%%%%%%%%%%%%%%%%%%%%%%%%%%%%%%%%
\subsection{Terminating states}
%%%%%%%%%%%%%%%%%%%%%%%%%%%%%%%%%%%%%%%%%%%%%%%%%%%

   It was shown in Refs.\ \cite{TWH.98,SatW.00} that the pair scattering from 
the $d_{3/2}$ and $f_{7/2}$ orbits into the aligned $g_{7/2}$ and $f_{7/2}$ 
orbits, which is entirely due to $t=0$ $np-$pairing, triggers the onset of 
collectivity for the states higher than $I=16^+$ in $^{48}$Cr. This can enhance 
the E2-transition rates between the yrast states with $I\geq 16^+$. This 
scenario is different from the standard one obtainable, for example, in 
cranked Nilsson-Strutinsky approach\cite{PhysRep}.  However, no experimental 
data on the states above $I=16^+$ are available in $^{48}$Cr so far.

  Theoretical analysis of the energy differences between terminating $f_{7/2}^n$ 
and $f_{7/2}^{n+1} d_{3/2}^{-1}$ states in the $A\sim 44$ nuclei within the 
Skyrme DFT showed that there is a good agreement with experiment for $N > Z$ nuclei 
and visible discrepancies for the $N=Z$ nuclei \cite{SS.06}. It was suggested
in Ref.\ \cite{SS.06} that the deviations from the data for the $N=Z$ nuclei are 
due to the $t=0$ $np-$pairing. However, isospin symmetry restoration is important 
for DFT description of the $N=Z$ nuclei and its inclusion improves the 
description of the data \cite{SDNR.10}. In addition, the DFT results sensitively
depends on the employed parametrization \cite{SDNR.10}.

%%%%%%%%%%%%%%%%%%%%%%%%%%%%%%%%%%%%%%%%%%%%%%%%%%%%%%%%%%%%%%%%%%%%%%%%%%%%%%%%
%%%%%%%%%%%%%%%%%%%%%%%%%%%%%%%%%%%%%%%%%%%%%%%%%%%%%%%%%%%%%%%%%%%%%%%%%%%%%%%%%%%%

%%%%%%%%%%%%%%%%%%%%%%%
\section{Conclusions}
%%%%%%%%%%%%%%%%%%%%%%%

  The physics of isoscalar and isovector neutron-proton pairing has 
been systematically reviewed in this article. At present, the existence 
of isovector $np$-pairing is well established. The isovector $np-$pairing 
is absolutely necessary in order to restore the isospin symmetry of the 
total wave function. Its strength is well defined by the isospin symmetry. A 
number of experimental observables such as binding energies of the $T=0$ and 
$T=1$ states in even-even and odd-odd $N=Z$ nuclei, the structure of rotational 
bands in $^{74}$Rb and pairing vibrations around $^{56}$Ni strongly support
its existence.

 On the contrary, the observed consequences of the $t=0$ $np-$pairing still 
remain illusive. The existence of the pair condensate in this channel 
sensitively depends on employed pairing strength. However, microscopic 
theories give no guidance on what strength has to be used for isoscalar 
$np-$pairing in the MF/DFT models. The use of experimental Wigner energies
as a tool to extract this strenghts faces the dilemma that these energies 
are not necessary due to isoscalar $np-$pairing. Other observables in 
non-rotating nuclei either do not support the existence of this type of 
pairing or insensitive to it. The systematic analysis of the rotational 
response of  $N\approx Z$ nuclei agrees with the picture which does not
involve isoscalar $np-$pairing. According to it (isovector mean-field theory),
at low spin, an isoscalar $np-$pair field is absent while a strong isovector 
pair field exists, which includes a large $np$ component, whose strength is 
determined by isospin conservation. Like in nuclei away from the $N = Z$ 
line, this isovector pair field is destroyed by rotation. In this high-spin 
regime, calculations without pairing describe accurately the data, provided 
that the shape changes and band termination are taken into account.
Although the current analysis does not support the existence of isoscalar
$np-$pairing, the possibility of its existence cannot be completely ruled 
out due to the limitations of existing  theoretical tools.

%%%%%%%%%%%%%%%%%%%%%%%%%%%%%%%%%%%%%%%%%%%%%%%
\section*{Acknowledgements}
%%%%%%%%%%%%%%%%%%%%%%%%%%%%%%%%%%%%%%%%%%%%%%%

This work has been supported by the U.S. Department of Energy under
the grant DE-FG02-07ER41459. Useful discussions with S. Frauendorf
are greatly appreciated.

\end{document}